\documentclass[twocolumn,showpacs,preprintnumbers,amsmath,amssymb]{revtex4}
\usepackage{graphicx}
\usepackage{dcolumn}
\usepackage{bm}
\usepackage{comment}
\usepackage{rotating}
\usepackage{longtable}
\usepackage{float}
\usepackage{eucal}
\usepackage{csquotes} 

\begin{document}

\title{Possibility of shears mechanism due to attractive interaction in the island of $A$ $\sim$ 140 region}

\author{S. Rajbanshi}

\affiliation{Dum Dum Motijheel College, Kolkata 700074, India}

\date{\today}

\begin{abstract}

The present work identify an island in the values of dimensionless parameter $\chi$ as +0.10 $<$ $\chi$ $<$ +0.55 outside which no magnetic rotational (MR) bands due to the repulsive interaction would be observed. The semi-classical calculations within the hybrid shears mechanism with the principal axis cranking model (HSPAC) reveal the possibility of the shears band due to the attractive interaction for nuclei in A $\sim$ 140 region. Though the limited information exists in literature, the HSPAC calculation exhibits the dipole bands above the 4389-keV and 4160-keV excited states in $^{145}$Sm and $^{146}$Eu nuclei, respectively, demand their candidature of the attractive shears bands. Other interesting consequence of the HSPAC calculations is the increasing nature of the $B(M1)$ and $B(E2)$ transition strengths with spin unlike the decreasing nature of the transition strengths ($B(M1)$ and $B(E2)$ values) of the MR bands. Thus measurement of the transition probability has been required on urgent basis to validate the strong candidature of such an extreme excitation mechanism in $^{145}$Sm and $^{146}$Eu nuclei.

\end{abstract}


\maketitle

\section{INTRODUCTION}

The generation of the band-like structure in weakly deformed nuclei near the shell closure is a long standing debate. These nuclei, generally, do not able to acquire the asymmetry in mass distribution like mid-shell nuclei that break the rotational symmetry which results the exhibition of the $E2$ band structure in their excitation spectrum \cite{abohr, defor1, defor2}. The current distributions of the few nucleons (particles and/or holes) outside the core play crucial role of generating the angular momentum in these nuclei. In fact, several exotic mechanisms have been observed due to the interplay between the holes and particles outside the core of the weakly deformed nuclei near shell closure \cite{frauen1}.

It is well established \cite{ajsim, hubel} that the \enquote{shears mechanism} is a general phenomenon of generating the angular momentum in weakly deformed nuclei, possess of few particles and holes orbiting around the core nucleus. The manifestation of the shears mechanism in terms of the observation of the band-like structure consisting of the strong $M1$ transitions called \enquote{magnetic rotation} (MR) has been found in the level structure of these nuclei \cite{amita}. The weak or unobserved cross-over $E2$ transitions provides strong support of the weakly deformed mass distribution to these nuclei. The repulsive interaction potential between the particles and holes favours the perpendicular coupling of the angular momentum vectors produced by them at the bandhead. The higher excited states of the MR band are generated by gradual closing of the angular momentum vectors. The MR band terminates when two blades are fully aligned giving maximum possible spin for the configuration of interest. At the end of the MR band, the current distribution becomes symmetric with respect to the total angular momentum vector ($I$).

It is noteworthy that the polarization of the core due to the occupancy of the few particles and holes in the high-$j$ shape driving orbitals plays a crucial role for generation of the MR band. This produces a certain core deformation which aligns the particles and holes along the respective axes, thus, make the nucleus favourable for the shears mechanism. The resulting core deformation, strongly correlated with the number of particle-hole pair of the MR band, can evolve throughout the band. Such interplay between the core polarization and shears mechanism has been observed in several weakly nuclei in different mass region. In shears mechanism, the core deformation plays a role similar to the reagent in chemical reaction but, here, both of these excitation's diminish one another. Therefore, a region of overlap of the core deformation and the number of particle-hole pair exists otherwise shears mechanism is forbidden for any other possible combinations. Introducing a dimensionless parameter ($\chi$) in the shears mechanism Clark and Macchiavelli has been explored such a region for the symmetric shears \cite{rmcla1}. It has been concluded then that shears mechanism is favourable for small values of $\chi$ whereas core rotation becomes more economical if $\chi$ is large ($>$ 50\%). Though the large number of the asymmetric shears (unequal shears blades $j_{p}$ and $j_{n}$ produced by the particles and holes, respectively) have been observed the evidence of the perfect symmetric shears are yet not exists in literature, to my best knowledge.

To explore the intrinsic characters as well as interplay between the core rotation and the shears mechanism of the MR bands the shears mechanism with the principal axis cranking (SPAC) model has been identified as the powerful tool \cite{pasern, podsvi, pasern1, rajban1, rajban2, rajban, rajban3, rajban4}. The main motivation for present work is to investigate the favourable region of the asymmetric and symmetric shears within the hybrid shears mechanism with the principal axis cranking model (HSPAC) by introducing $\chi$ in the SPAC model for the weakly deformed nuclei near shell closure. The emphasis will be given to explore the possibility of the attractive shears (shears blades are produced by the similar type of the valence particles $i. e$ either particles or holes) of these nuclei as hinted by Clark and Macchiavelli. The possible candidature of the attractive shears for the $N$ = 83 $^{145}$Sm \cite{odhara} and $^{146}$Eu \cite{eidegu} isotones will be discussed.

\section{Framework of the model}

The theoretical calculations within the semi-classical shears model as well as the SPAC model have been performed to explore the intrinsic structural properties of the MR bands in $A$ $\sim$ 140 mass region \cite{pasern, podsvi, pasern1, rajban1, rajban, rajban3, rajban4}. In the SPAC model, the rotation of core has been taken into consideration for better understanding of the experimental results of the MR bands. Clark and Macchiavelli identified an island for shears mechanism by introducing a dimensionless parameter ($\chi$) in the symmetric shears model in which magnitude of the shears blades, $j_{p}$ and $j_{n}$ produced by the particles and holes, respectively, are equal. Since the perfect symmetric shears are yet remain to be unobserved the present work focus to construct a hybrid model by introducing the $\chi$ in the SPAC model to describe the observed features of the asymmetric shears bands. Due to this the present model has been termed as the hybrid shears mechanism with the principal axis cranking (HSPAC) model.

\begin{figure}[t]
\centering
\setlength{\unitlength}{0.05\textwidth}
\begin{picture}(10,7.0)
\put(-2.5,-3.0){\includegraphics[width=0.70\textwidth, angle = 0]{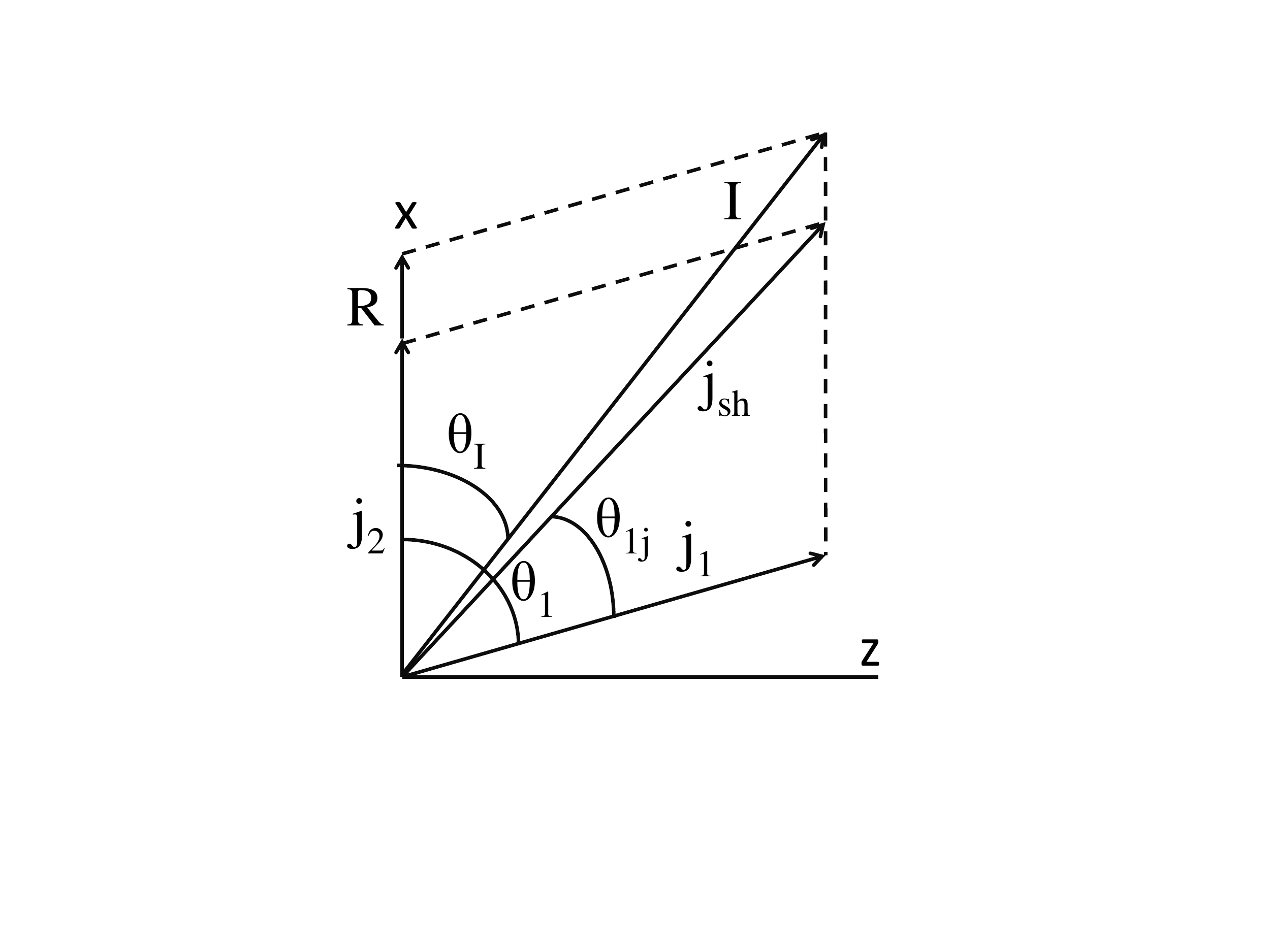}}
\end{picture}
\caption{\label{theo-cal} Schematic description of the angular momentum coupling scheme, responsible for repulsive interaction, used in the semiclassical HSPAC model calculation and the orientation of shears blades $j_{1}$ and $j_{2}$ for normal initial alignment.}
\end{figure}

In this model, total energy of an excited state $E(I)$ can be expressed as, 

\begin{equation}
{E(I)= \cfrac{R^{2}\left(\text{I, $\theta_{1}$}\right)}{2J(I)} + v_{2}P_{2}(cos(\theta_{1}))}
\label{spac1}
\end{equation}

where, first and second terms represent the collective and quasi-particle (shears) interaction energy contribution. Here, $\theta_{1}$ is the angle of  $\overrightarrow{j_{1}}$ with respect to the rotational axis ($\overrightarrow{R}$) whereas the direction of $\overrightarrow{j_{2}}$ is set along the rotational axis ($\overrightarrow{R}$) [as, shown in Fig. \ref{theo-cal}]. 

The spin of the state ($I$) is thus related to the $\overrightarrow{j_{1}}$, $\overrightarrow{j_{2}}$ and $\overrightarrow{R}$ as,

$\overrightarrow{R}$ = $\overrightarrow{I}$ - $\overrightarrow{j_{1}}$ - $\overrightarrow{j_{2}}$

or, $R^{2}$ = $I^{2}$ + $j_{1}^{2}$ + a.$j_{1}^{2}$ - 2$I$$j_{1}$(a.cos$\theta_{I}$ + cos($\theta_{1}$ - $\theta_{I}$)) + a.$j_{1}^{2}$.cos$\theta_{1}$

Expressing $E(I)$ in the reduced form $\hat{E}(\hat{I})$ = $E(I)$$\times$$\cfrac{J(I)}{2j_{1}^{2}}$, the equation \ref{spac1} becomes,

\begin{center}
$\hat{E}(\hat{I})= \hat{I}^{2} + \cfrac{1 + a^{2}}{4} -  \cfrac{1}{2}(\sqrt{4\hat{I}^{2} - sin^{2}\theta_{1}} (a + cos\theta_{1}) + $
\end{center}
\vskip -0,7cm
\begin{equation}
sin^{2}\theta_{1}) + \cfrac{a}{2}cos\theta_{1} + \cfrac{\chi}{4} cos^{2}\theta_{1} - \cfrac{\chi}{12}
\label{spac2}
\end{equation}

where, $\chi$ = $\cfrac{ J(I)}{j_{1}^2/3v_{2}}$ is a dimensionless quantity, determines the contribution of the core rotation in the shears band. The sign of the $\chi$ represents the nature of the potential i. e. attractive ($\chi$ = -ve) or repulsive ($\chi$ = +ve) interaction.

For each value of the reduced spin $\hat{I}$, the angle $\theta_{1}$ between the shears blades $j_{1}$ and $j_{2}$ (= a$j_{1}$) can be obtained by  minimized $\hat{E}(\hat{I})$ w.r.t. the angle $\theta_{1}$, 

\begin{equation}
\cfrac{\partial \hat{E}(\hat{I},\theta_{1})}{\partial \theta_1}=0.
\label{spac3}
\end{equation}

Simplification of this gives the relation between $\hat{I}$ and $\theta_{1}$ as follows,

\begin{equation}
\sqrt{4\hat{I}^{2} - sin^{2}\theta_{1}} +  \cfrac{cos\theta_{1} (a + cos\theta_{1})}{\sqrt{4\hat{I}^{2} - sin^{2}\theta_{1}}} -(2+\chi)cos\theta_{1} - a = 0 
\label{spac4}
\end{equation}

Thus, the rotational frequency ($\hat{\omega}$) for the state with reduced spin $\hat{I}$ can be obtained from,

\begin{equation}
\hat{\omega} = 2\hat{I} (1 - \chi + \cfrac{\sqrt{4\hat{I}^{2} - sin^{2}\theta_{1}} - a}{cos\theta_{1}}) 
\label{spac5}
\end{equation}

This minimized value of $\theta_{1}$ is then used to determine the $B(M1)$ and $B(E2)$ transition rates from the following relations,

\begin{equation}
B(M1) = \cfrac{3}{8\pi}\left[\text{$g_{1}^{*}j_{1}$sin($\theta_{1}$ - $\theta_{I}$) - $g_{2}^{*}j_{2}$sin($\theta_{I}$)}\right]^{2}, \text{and}
\label{spac6}
\end{equation}

\begin{equation}
B(E2) = \cfrac{15}{128\pi}\left[\text{Q$_{eff}$sin$^{2}$($\theta_{1j}$) + Q$_{coll}$cos$^{2}$($\theta_{I}$)}\right]^{2},
\label{spac7}
\end{equation}

where, $g_{1}^{*}$ = $g_{1}$ - $g_{R}$, $g_{2}^{*}$ = $g_{2}$ - $g_{R}$ and $g_{R}$ = Z/A. Here, Q$_{eff}$ and Q$_{coll}$ are the quasi-particle and collective quadrupole moments, respectively.

The effectiveness of this model over SPAC model is the reduction of the free parameters. In SPAC model, two independent parameters the core moment of inertia ($J(I)$) and the particle-hole interaction potential ($v_{2}$) are defined to characterize the core rotation and the shears mechanism, respectively. In the present model, only one parameter $\chi$ is defined which is representative of contribution of the angular momentum produced by the rotation of the deformed core with respect to the shears mechanism for generation of the MR bands. The present model allows to defining a region in $\chi$ outside which rotation of core dominates over the shears mechanism, therefore, shears mechanism ceased. For small values of the $\chi$ ($<$ 0.5) the shears mechanism dominates while for higher values of $\chi$ ($>$ 0.5) the core rotation plays the important role for generation of the MR band.

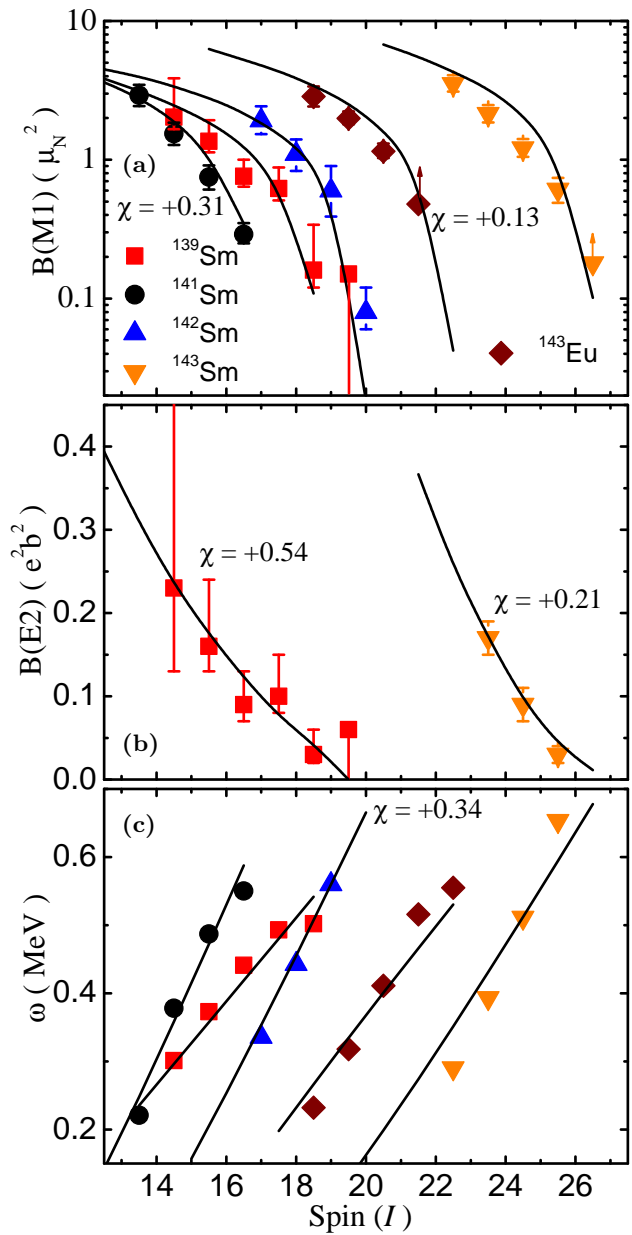
\begin{figure}[t]
\centering
\setlength{\unitlength}{0.05\textwidth}
\begin{picture}(10,18.0)
\put(-0.6,-0.3){\includegraphics[width=0.57\textwidth, angle = 0]{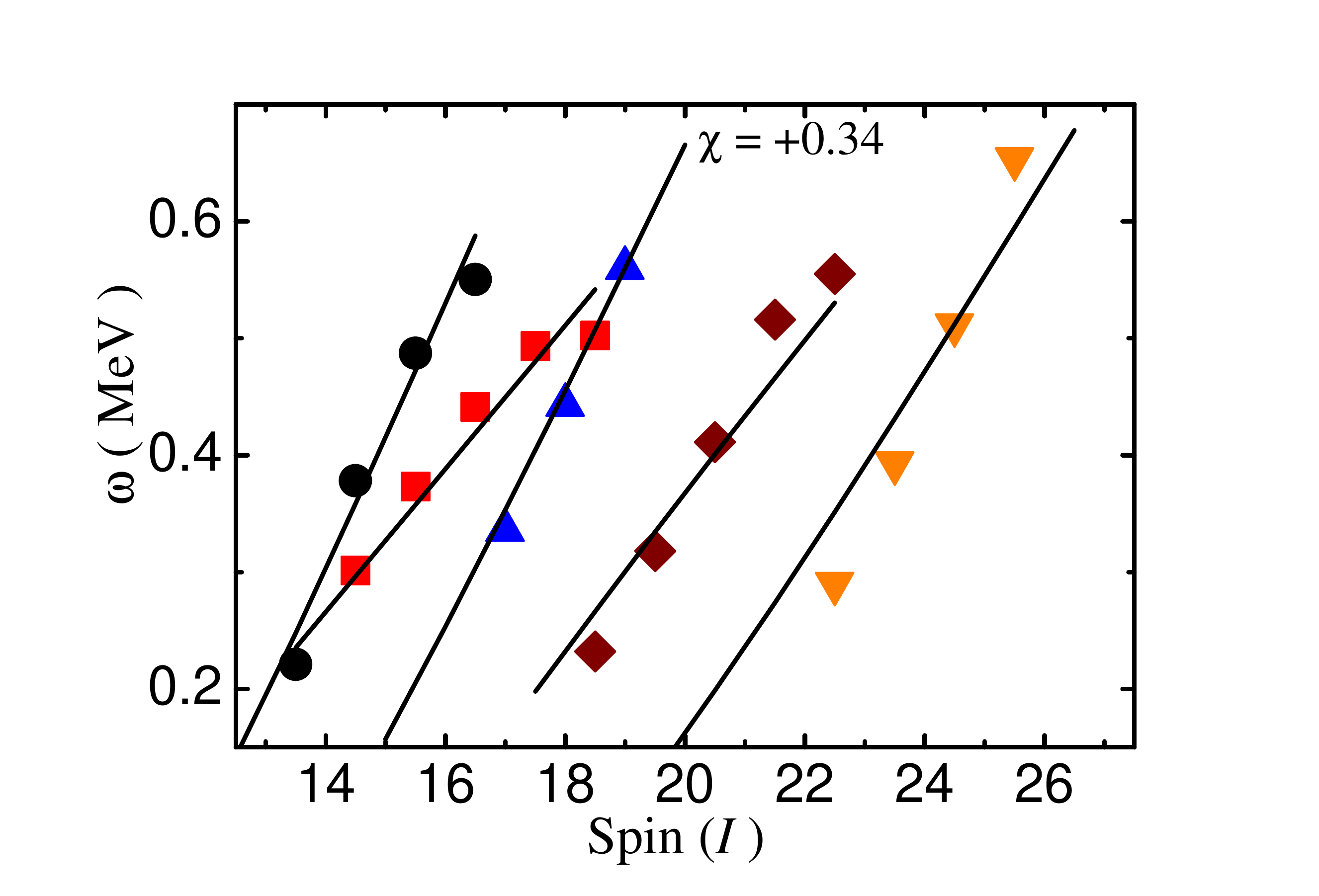}}
\put(-0.6,+5.4){\includegraphics[width=0.57\textwidth, angle = 0]{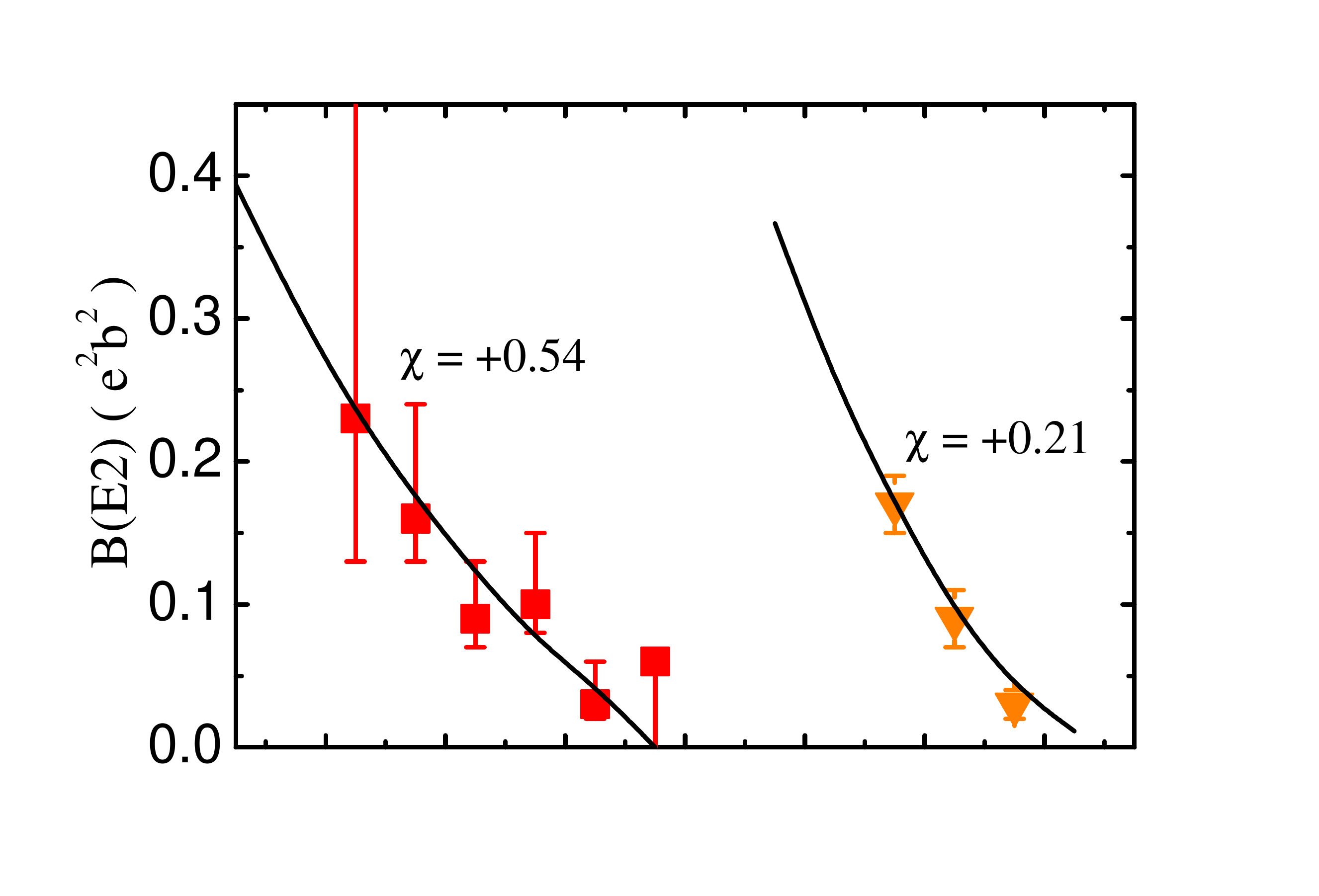}}
\put(-0.6,+11.1){\includegraphics[width=0.57\textwidth, angle = 0]{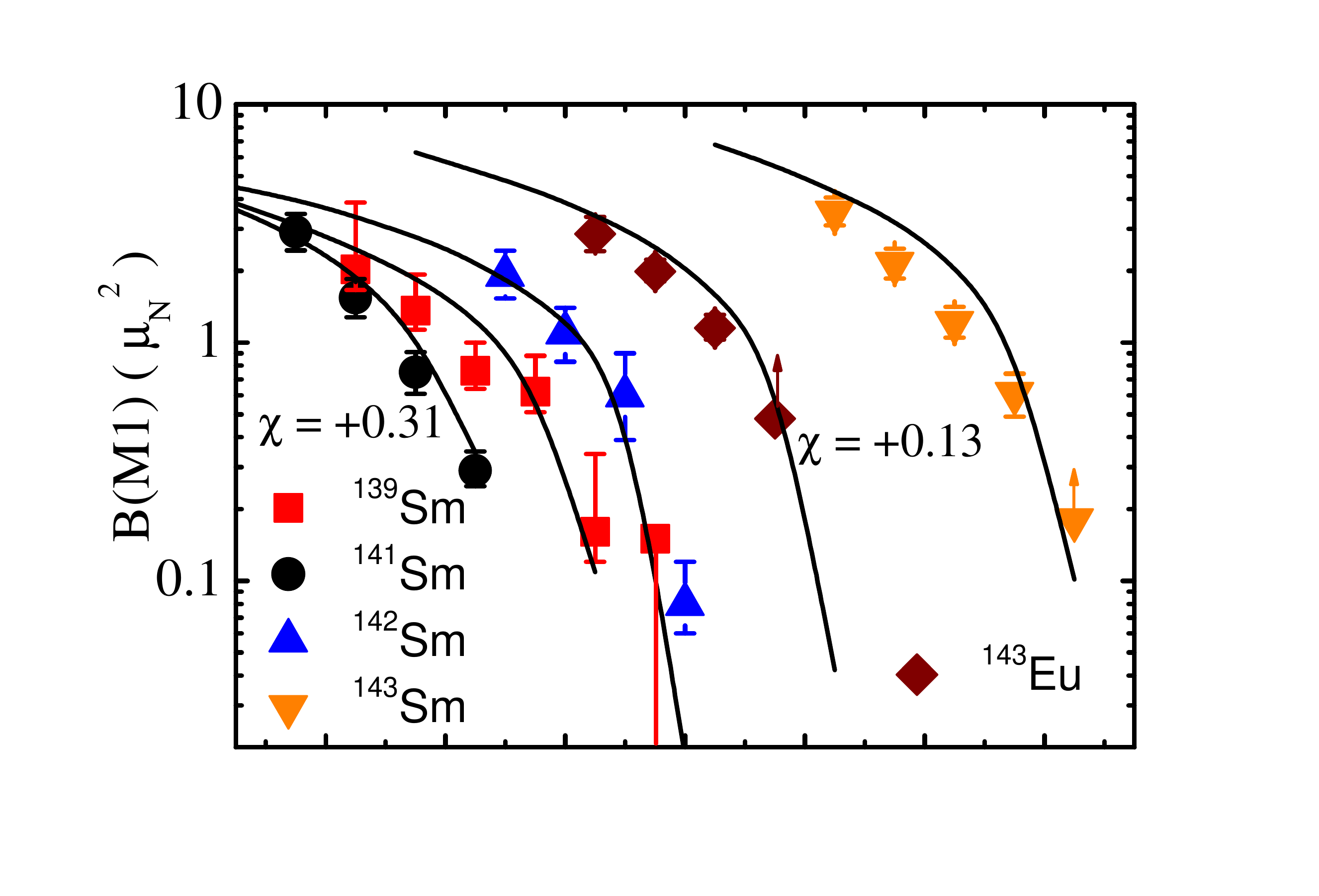}}
\put(1.7,15.7){\textbf{(a)}}
\put(1.7,7.1){\textbf{(b)}}
\put(1.7,5.9){\textbf{(c)}}
\end{picture}
\caption{\label{theo-cal2} (Color  online) The comparison of the experimental (a) $B(M1)$ (b) $B(E2)$ and (c) $\omega$ values with the HSPAC model calculations (solid lines) for the MR bands in $^{139, 141-143}$Sm and $^{143}$Eu nuclei in mass $A$ $\sim$ 140 region. The experimental results are adopted from the earlier works on the respective nuclei \cite{pasern, podsvi, pasern1, rajban1, rajban, rajban3, rajban4}.}
\end{figure}

\section{Calculation for the $\bf{Sm}$ and $\bf{Eu}$ nuclei}

A large number of dipole bands have been observed in the $A \sim 140$ mass region. The dipole bands of the $^{139,141,142, 143}$Sm, $^{141,143}$Eu and $^{142}$Gd nuclei in this mass region have been identified as MR bands from the experimental measurements of the $B(M1)$ values along with the theoretical TAC and SPAC model calculations \cite{pasern, podsvi, pasern1, rajban1, rajban, rajban3, rajban4}. In the $A$ = 140 mass region, the valence protons and neutrons quasi-particles in the $g_{7/2}$/$d_{5/2}$ and $h_{11/2}$ orbitals, respectively, make the nuclear system preferable for generating the MR bands. The assigned configurations of these MR bands along with the obtained equilibrium deformation parameters have been listed in Table \ref{tactrs}. The tilted axis cranking model and total routhian surface calculations in most of the cases are in agreement to each other. This allows us to unwavering the shape of the nucleus associated with the assumed configuration. The calculations show that the deformation increases as neutron number decreases, as for example, the deformation associated with the MR bands in $N = 78$ nuclei is much larger (by a factor of two) than the deformation of the MR bands in $N = 80$ nuclei (Table \ref{tactrs}).

To get better understanding of the effect of the core rotation on the MR bands in this mass region we have performed the Hybrid shears mechanism with the principal axis cranking (HSPAC) model calculations. In this calculation, the configurations of the MR bands have been adopted from the Ref. \cite{pasern, podsvi, rajban1, rajban, rajban3, rajban4} as given in Table \ref{tactrs}. The parameters used in the calculations are similar to those of the SPAC calculations of the respective nuclei. Using the same sets of parameters as of the SPAC calculations it has been observed that the experimental results are in well agreement within the present HSPAC calculations. The experimental rotational frequency ($\omega$), $B(M1)$ and $B(E2)$ values (taken from the previous works \cite{pasern, podsvi, rajban1, rajban, rajban3, rajban4}) of these bands, as shown in Fig. \ref{theo-cal2}, are in good agreement with calculations performed for the values of the parameter $\chi$ as +0.21, +0.34, +0.31 and +0.54, for the Sm isotopes having $N$ = 81, 80, 79 and 77, respectively. It has been observed that the value of $\chi$ increases from the +0.21 for $N$ = 81 to +0.54 for $N$ = 77 isotopes of Sm. The same calculations has been performed for the Eu isotopes. Though the observation of the MR bands are limited only to the $^{141}$Eu ($N$ = 78) and $^{143}$Eu ($N$ = 80) nuclei the calculations shows that core rotation plays significant role in $^{141}$Eu ($\chi$ = +0.41)  whereas shears mechanism dominates for $^{143}$Eu ($\chi$ = +0.13). This reflects the fact that the effect of the core rotation on the MR bands increases as one goes away from the shell closure. It is in-coincidence with our prime knowledge of the deviation of the nuclear shape from spherical becomes more prominent for nuclei close to the mid-shell.

\begin{table*}[t]
\caption{\label{tactrs} The proposed configurations  and the associated equilibrium deformation parameters for the MR bands in mass $A = 140$ region.}
\begin{ruledtabular}
\begin{tabular}{cccccccc}
Nucleus & Configuration &  ($\epsilon_{2}$, $\gamma$) [TAC]\footnotemark[1] & ($\beta$, $\gamma$) [TRS] & Shape & $j_{1}$ & $j_{2}$ & $\chi$\\
\hline\hline

$^{141}$Eu DB1 ($N = 78$) & $\pi h_{11/2}^{1} {\otimes}$ $\nu{h}_{11/2}^{-2}$ & -0.07, 0$^\circ$ & 0.16, -87$^\circ$ & Oblate & 5.5$\hbar$ & 8$\hbar$  & +0.41\\

$^{143}$Eu DB I ($N = 80$) & $\pi h_{11/2}^{2} {\otimes}$ $\nu{h}_{11/2}^{-2}$ $\pi{g}_{7/2}^{-1}$ & & 0.07, -10$^\circ$ & Prolate & 11.5$\hbar$ & 8$\hbar$  & +0.13\\

$^{139}$Sm ($N = 77$) & $\pi h_{11/2}^{2} {\otimes}$ $\nu{h}_{11/2}^{-1}$ & +0.10, 0$^\circ$ & 0.18, 30$^\circ$ & Prolate  & 5.5$\hbar$ & 9$\hbar$ & +0.54 \\

$^{141}$Sm ($N = 79$) & $\pi h_{11/2}^{2} {\otimes}$ $\nu{h}_{11/2}^{-1}$ & & 0.16, 37$^\circ$& Prolate & 5.5$\hbar$  & 9$\hbar$ &+0.31\\

$^{142}$Sm ($N = 80$) & $\pi h_{11/2}^{1} {\otimes}$ $\nu{h}_{11/2}^{-2}$ $\pi{g}_{7/2}^{-1}$ & & 0.07, -94$^\circ$ & Oblate & 5.5$\hbar$ & 11.5$\hbar$ & +0.34\\

$^{143}$Sm ($N = 81$) & $\pi h_{11/2}^{4} {\otimes}$ $\nu{h}_{11/2}^{-1}$ $\pi{g}_{7/2}^{-2}$ & & 0.10, 0$^\circ$ & Prolate & 9.5$\hbar$ & 14$\hbar$ & +0.21\\

\end{tabular}
\end{ruledtabular}
\footnotetext[1]{The minimum values of the deformation parameters ($\epsilon_{2}$, $\gamma$) for $^{141}$Eu and $^{139}$Sm are taken from the Refs. \cite{podsvi} and \cite{pasern}, respectively. The deformation parameters ($\beta$, $\gamma$) are either taken from the previous Refs. \cite{rajban1, rajban, rajban3, rajban4} or obtain using the prescription as described in Refs. \cite{naza1, naza2}.}

\end{table*}

\begin{figure}[b]
\centering
\setlength{\unitlength}{0.05\textwidth}
\begin{picture}(10,13.0)
\put(-0.5,-0.8){\includegraphics[width=0.53\textwidth, angle = 0]{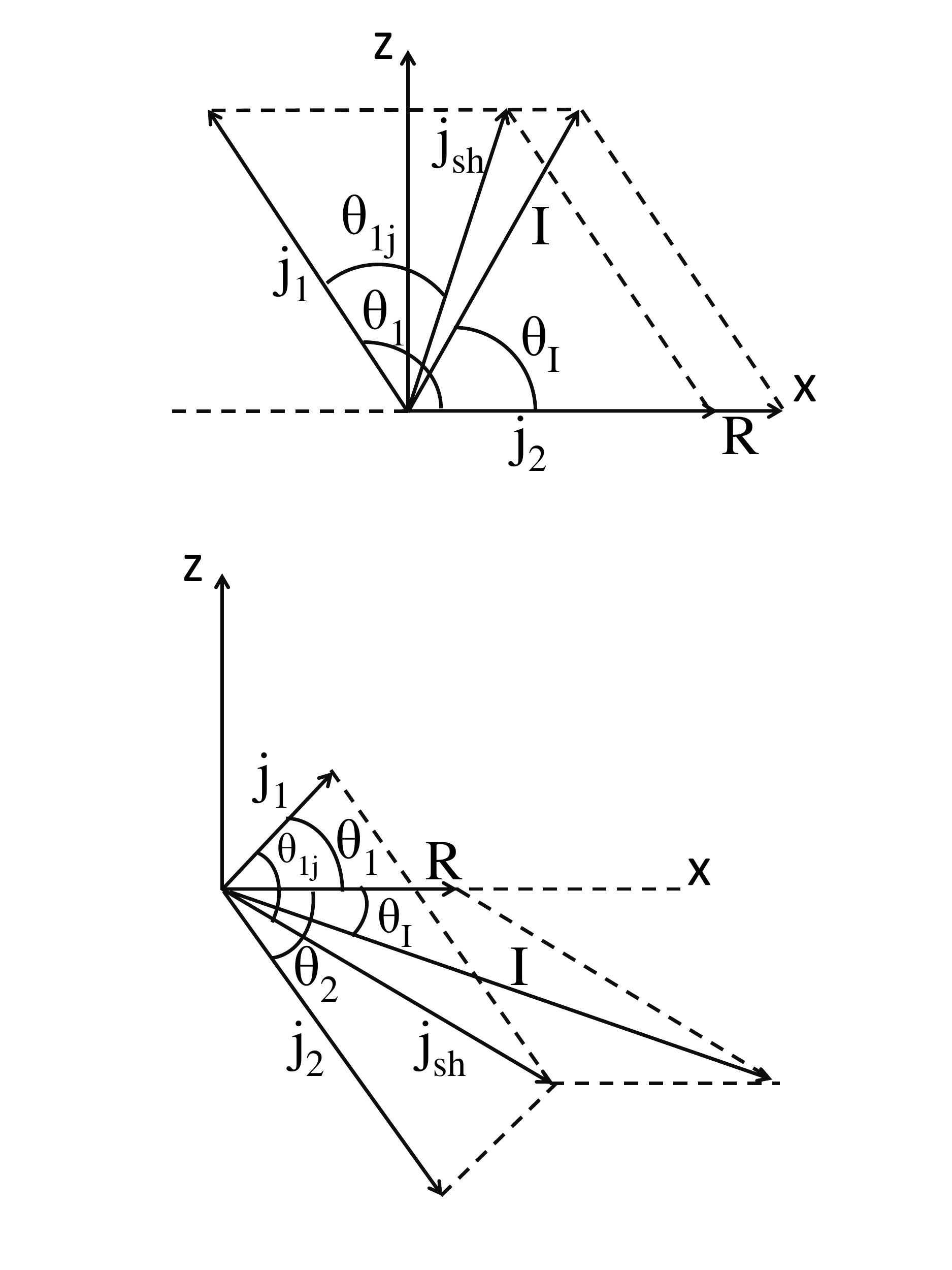}}
\put(0.7,10.5){\textbf{(a)}}
\put(0.7,1.2){\textbf{(b)}}
\end{picture}
\caption{\label{theo-cal3} The angular momentum vector coupling scheme for the attractive interaction potential between the valency particles. Two possible configuration emerge out due to the orientation of the angular momentum vectors are (a) along the rotational axis and anti-parallel to it (b) perpendicular to the rotational axis.}
\end{figure}

The semi-classical calculations within the HSPAC model show that the large collective contribution ($\chi$ = +0.54) of the MR in $^{139}$Sm (N = 77) is in agreement with the TAC and the SPAC results as performed in the previous work. On the other hand, least rotational contribution ($\chi$ = +0.13) results the maximal contribution of the angular momentum arises from the quasi-particle interaction for the MR band in $^{143}$Eu. All the observed MR bands in the $A$ $\sim$ 140 region are fall in between these two extreme values of $\chi$. Thus, one may identify an island in the values of $\chi$ as +0.10 $<$ $\chi$ $<$ +0.55 outside which no MR bands due to the repulsive interaction would be observed.

\section{Possibility of the MR bands due to the attractive interaction}

Until today, the observation of the MR bands in the weakly deformed nuclei is restricted to only due to the repulsive interaction. Clark and Macchiavelli first explain the possibility of the shears band due to the attractive interaction of the similar type of the valence particles $i. e$ either particles or holes.

Depending of the deformation of the nuclear system alignment of the anti-aligned valency particles two distinct situation would occurs. The first one is the alignment of the particle(holes) along the rotational axis (x) and anti-parallel to it. In this case, with increasing frequency the angular momentum vector anti-aligned to the rotational axis tries to align along the rotational axis. The energy of this configuration, as shown in Fig. \ref{theo-cal3} (a), can be represented by Eq. \ref{spac2} considering the sign of $\chi$ as negative. Consequently, the shears frequency and the $B(M1)$ and $B(E2)$ transitions strength can be calculated following the Eqs. \ref{spac5}, \ref{spac6} and \ref{spac7}, respectively.

\begin{figure}[b]
\centering
\setlength{\unitlength}{0.05\textwidth}
\begin{picture}(10,12.5)
\put(-0.5,-0.2){\includegraphics[width=0.50\textwidth, angle = 0]{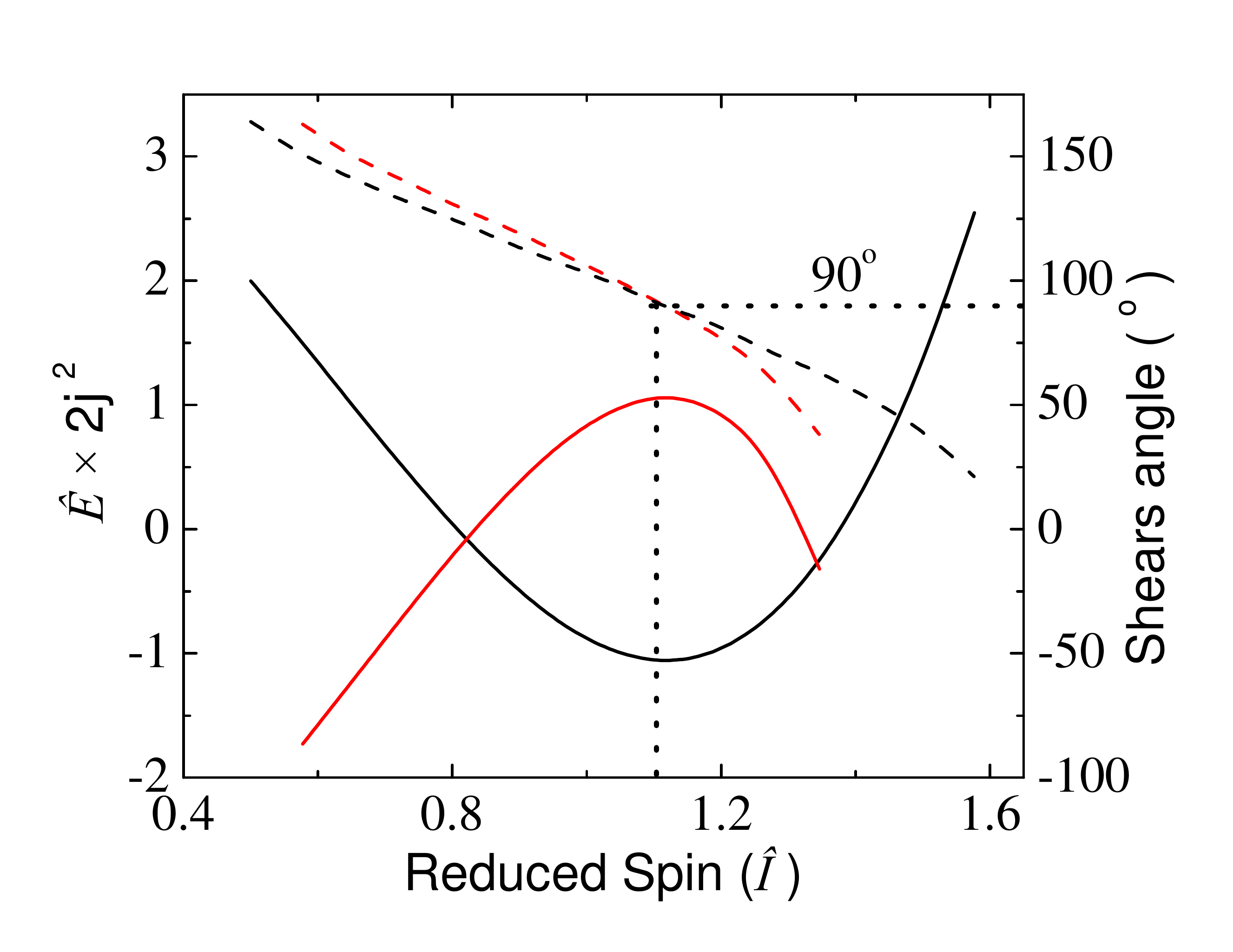}}
\put(-0.83,5.6){\includegraphics[width=0.50\textwidth, angle = 0]{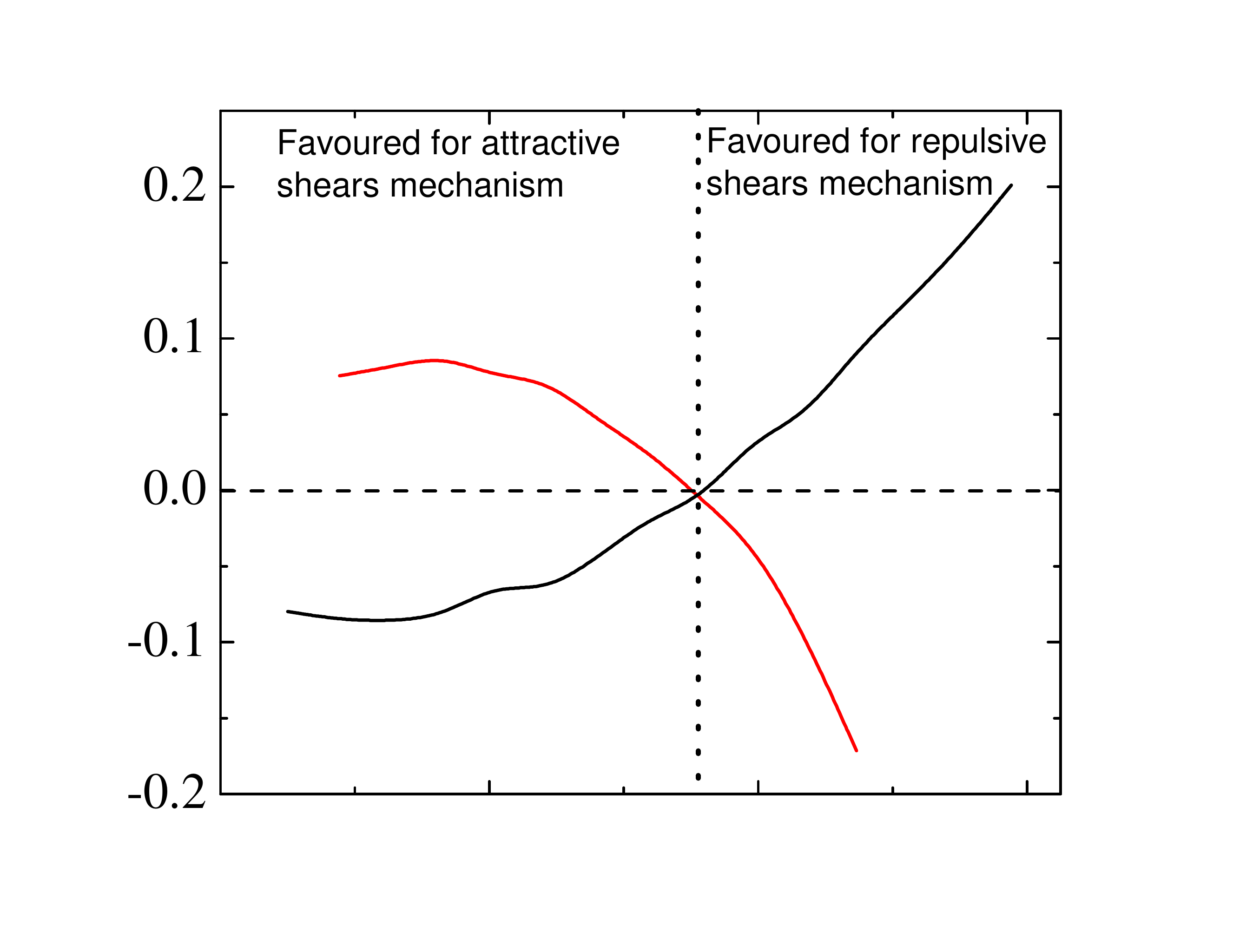}}
\put(1.5,11.3){\textbf{(a)}}
\put(1.5,2.5){\textbf{(b)}}
\put(-0.0,9.9){\rotatebox{90}{\textbf{$\hat\omega$}}}
\end{picture}
\caption{\label{theo-cal4} (Color  online) The calculated (a) shears frequency ($\hat\omega$) and (b) energy (solid lines) and the shears angle (dash lines) within the HSPAC model considering the attractive (red lines) and repulsive (black lines) interaction potentials with the values of $\chi$ as -0.15 and +0.15, respectively. In the calculation the values of $j_{1}$ and a are adopted as 6.5 and 2, respectively.}
\end{figure}

\begin{figure}[b]
\centering
\setlength{\unitlength}{0.05\textwidth}
\begin{picture}(10,7.1)
\put(-0.5,-0.2){\includegraphics[width=0.52\textwidth, angle = 0]{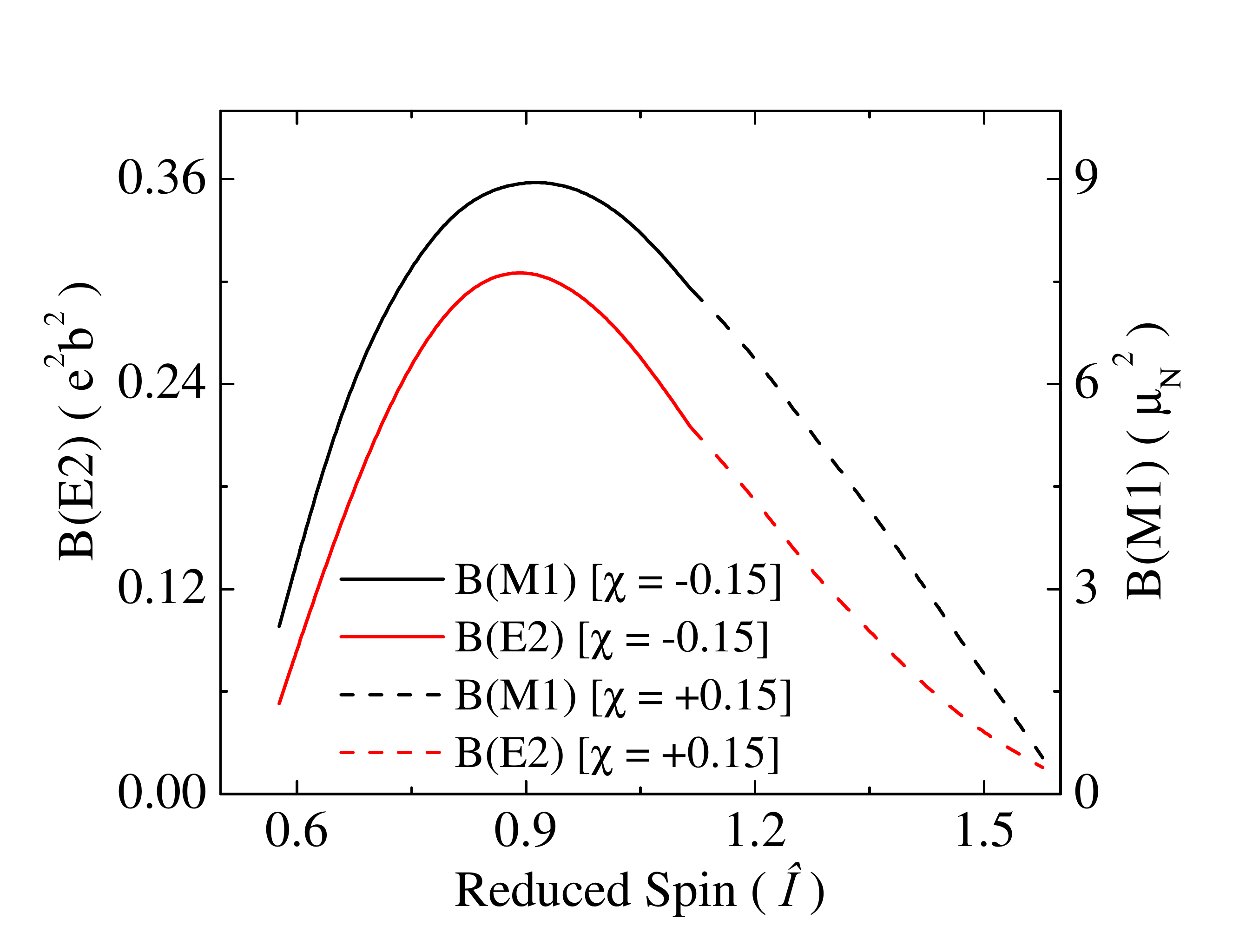}}
\end{picture}
\caption{\label{theo-cal5} (Color  online) The calculated $B(M1)$ (black lines) and $B(E2)$ (red lines) values within the HSPAC model considering the attractive (solid lines) and repulsive (dash lines) interaction potentials with the values of $\chi$ as -0.15 and +0.15, respectively. In the calculation the values of the $Q_{eff}$, $Q_{coll}$, $g_{1}$ and $g_{2}$ are 2.5 eb, 0.5 eb, -0.15 and +1.21, respectively. The other parameters are exactly same as given in the caption of Fig. \ref{theo-cal4}.}
\end{figure}

On the basis of the above formalism some distinct characteristic features of the shears bands due to the attractive interaction ($\chi$ = -ve) have been found in comparison with the well-established MR bands due to the repulsive interaction potential as follows.

i. Unlike the smoothly increasing nature of the shears frequency in the MR bands the calculated frequency shows a decreasing trend with spin when the attractive interaction has been considered as shown in Fig. \ref{theo-cal4} (a).

ii. The attractive interaction potential produces minimum energy for the anti-parallel coupling of the angular momentum vectors at the bandhead whereas the perpendicular coupling is favourable for the repulsive potential (conventional MR bands) as depicted in Fig. \ref{theo-cal4} (b).

iii. The conventional MR bands, produce due to the repulsive interaction, terminate when the shears blades are fully aligned i. e. the angle between them is zero but in present case of the attractive potential the perpendicular coupling of the blades gives maximum energy, therefore, band ends at this point (Fig. \ref{theo-cal4} (b)) .

iv. The B(M1) and B(E2) transition probabilities show a sharp increasing nature with spin of the states in contrast to the decreasing behaviour of the same for the conventional MR bands (Fig. \ref{theo-cal5}).

Another situation would occurs, as described in Fig. \ref{theo-cal3} (b), when the valency particles favour their angular momentum vectors along the deformation (symmetry) axis and anti-parallel to it. The higher energy states within this configuration have been generated by the simultaneous closing of the anti-aligned vectors towards the rotational axis. The energy of the state $I$ within this configuration can be expressed as

\begin{equation}
{E(I)= \cfrac{R^{2}\left(\text{I, $\theta_{1}$, $\theta_{2}$}\right)}{2J(I)} + v_{2}P_{2}(cos(\theta_{1} + \theta_{2}))},
\label{spac8}
\end{equation}

where, first and second terms represent the collective and quasi-particle (shears) interaction energy contribution. Here, $\theta_{1}$ and $\theta_{2}$ are the angle of  $\overrightarrow{j_{1}}$ and $\overrightarrow{j_{2}}$ with respect to the rotational axis ($\overrightarrow{R}$), respectively, as shown in Fig. \ref{theo-cal3} (b). Using the dimensionless parameters $\hat{I}$, a (=$j_{2}$/$j_{1}$) and $\chi$ (as defined earlier) the energy of the state reduced to $\hat{E}$ as

\begin{center}
$\hat{E}(\hat{I})= \hat{I}^{2} + \cfrac{1 + a^{2}}{4} -  \cfrac{1}{2}(\sqrt{4\hat{I}^{2} - X^{2} sin^{2}(\theta_{1j} - \theta_{1})}   (cos\theta_{1} + a \times cos\theta_{2}) + X sin(\theta_{1j} - \theta_{1}) (a \times sin\theta_{2} - sin\theta_{1})) + $
\end{center}
\vskip -.7cm
\begin{equation}
\cfrac{a}{2}cos(\theta_{1} + \theta_{2}) + \cfrac{\chi}{4} cos^{2}(\theta_{1} + \theta_{2}) - \cfrac{\chi}{12}
\label{spac9}
\end{equation}

where, $X$ = $\sqrt{1 + a^{2} + 2a \times cos(\theta_{1} + \theta_{2})}$ is a dimensionless quantity and 

\begin{equation}
\theta_{1j} = tan^{-1} \cfrac{j_{2} sin(\theta_{1} + \theta_{2})} {j_{1} + j_{2}cos(\theta_{1} + \theta_{2})}
\label{spac10}
\end{equation}
 
is the angle of $\overrightarrow{j_{1}}$ with respect to the vector $\overrightarrow{j_{sh}}$ = $\overrightarrow{j_{1}}$ + $\overrightarrow{j_{2}}$. For each value of the reduced spin $\hat{I}$, the angles $\theta_{1}$ and $\theta_{2}$ can be obtained by  minimized $\hat{E}(\hat{I})$ w.r.t. the angles $\theta_{1}$ and $\theta_{2}$, 

\begin{equation}
\cfrac{\partial^{2} \hat{E}(\hat{I},\theta_{1},\theta_{2})}{\partial \theta_1 \partial \theta_2}=0.
\label{spac11}
\end{equation}

This 2-D minimization condition reduced to 1-D minimization condition when the variation of $\theta_{2}$ has been assumed as

\begin{equation}
\theta_{2} = \theta_{2}^{i} (\theta_{1}/\theta_{1}^{i})^\alpha,
\label{spac12}
\end{equation}

where, $\theta_{1}^{i}$ and $\theta_{2}^{i}$ are the initial values of the angles $\theta_{1}$ and $\theta_{2}$ and taken them as 90$^{\circ}$ in the present calculation. The quantity $\alpha$ represents how values of $\theta_{2}$ change with respect to the angle $\theta_{1}$. 

Thus, the rotational frequency ($\hat{\omega}$) for the state with reduced spin $\hat{I}$ can be obtained from,

\begin{equation}
\hat{\omega} = 2\hat{I} (1 - \cfrac{cos\theta_{1} + a \times cos\theta_{2}} {\sqrt{4\hat{I}^{2} - X^{2} sin^{2}(\theta_{1j} - \theta_{1})}} )  
\label{spac13}
\end{equation}

This minimized values of $\theta_{1}$ and corresponding values of $\theta_{2}$ are then used to determine the $B(M1)$ and $B(E2)$ transition rates from the following relations,

\begin{equation}
B(M1) = \cfrac{3}{8\pi}\left[\text{$g_{1}^{*}j_{1}$sin($\theta_{1}$ + $\theta_{I}$) - $g_{2}^{*}j_{2}$sin($\theta_{2}$ - $\theta_{I}$)}\right]^{2}, \text{and}
\label{spac14}
\end{equation}

\begin{equation}
B(E2) = \cfrac{15}{128\pi}\left[\text{Q$_{eff}$sin$^{2}$($\theta_{1j}$) + Q$_{coll}$cos$^{2}$($\theta_{I}$)}\right]^{2},
\label{spac15}
\end{equation}

where, $g_{1}^{*}$ = $g_{1}$ - $g_{R}$, $g_{2}^{*}$ = $g_{2}$ - $g_{R}$ and $g_{R}$ = Z/A. Here, Q$_{eff}$ and Q$_{coll}$ are the quasi-particle and collective quadrupole moments, respectively.

\begin{figure}[t]
\centering
\setlength{\unitlength}{0.05\textwidth}
\begin{picture}(10,7.5)
\put(-0.5,-0.2){\includegraphics[width=0.52\textwidth, angle = 0]{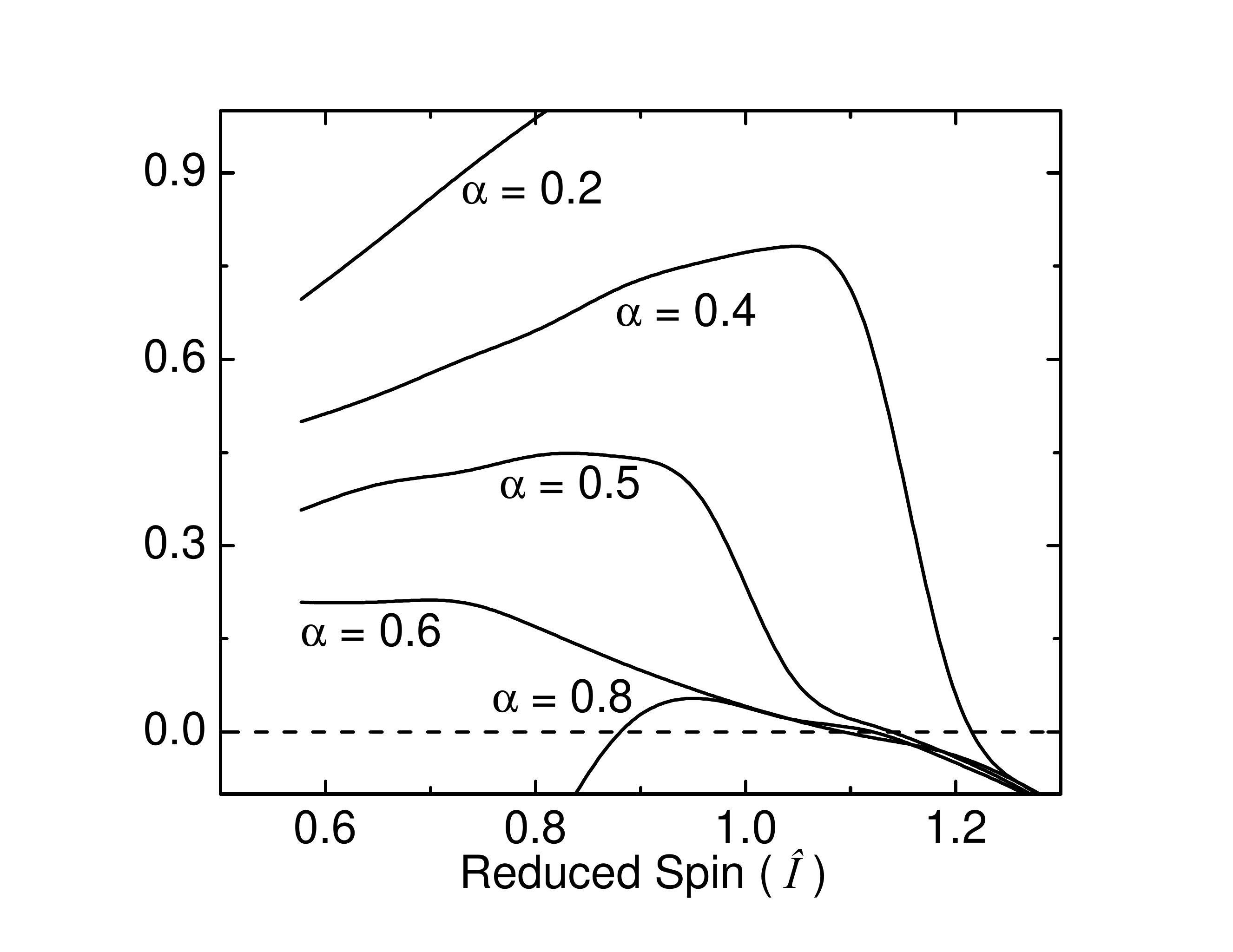}}

\put(1.6,6.0){\textbf{(a)}}

\put(+0.2,3.6){\rotatebox{90}{\textbf{$\hat\omega$}}}
\end{picture}
\caption{\label{theo-cal6} (Color  online) The calculated shears frequency ($\hat\omega$) within the HSPAC model considering the attractive interaction potential and initial anti-parallel alignment to the symmetry axis (Fig. \ref{theo-cal3} (b)) with the values of $\chi$ as -0.15. In the calculation the values of $j_{1}$ and a are adopted as 6.5 and 2, respectively.}
\end{figure}

The calculated results within the initial aligned and anti-aligned to the symmetry (deformation) axis are well reproduced the results as obtained from the rotation aligned consideration (Figs. \ref{theo-cal4} and \ref{theo-cal5}) except here the calculated frequency shows strong correlation with the parameter $\alpha$ (Fig. \ref{theo-cal6}). The decreasing features of $\hat\omega$ has been found for the values of $\alpha$ $>$ 0.5 and above this value of $\alpha$ the $\hat\omega$ start to exhibiting rapid increasing behaviour with $\hat{I}$. This result is not furious as for small values of $\alpha$, say $\alpha$ = 0.2, the angle $\theta_{2}$ remains almost constant, therefore the vector $\overrightarrow{j_{2}}$ directed along the symmetry axis favours such increasing trend of $\hat\omega$ with the $\hat{I}$ throughout the band.

\begin{figure}[b]
\centering
\setlength{\unitlength}{0.05\textwidth}
\begin{picture}(10,8.0)
\put(-4.0,-1.5){\includegraphics[width=0.59\textwidth, angle = 0]{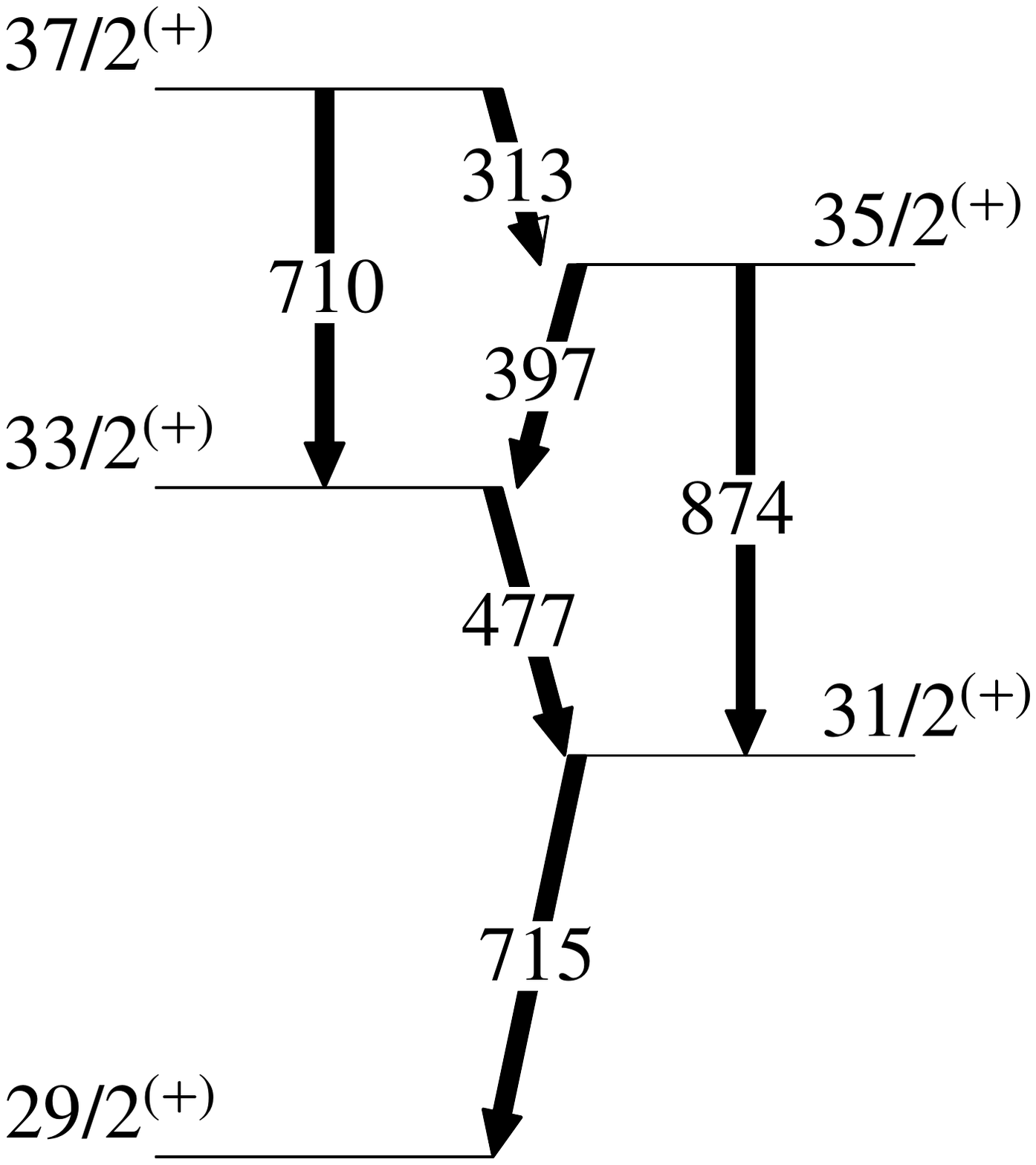}}
\put(+2.2,-1.2){\includegraphics[width=0.60\textwidth, angle = 0]{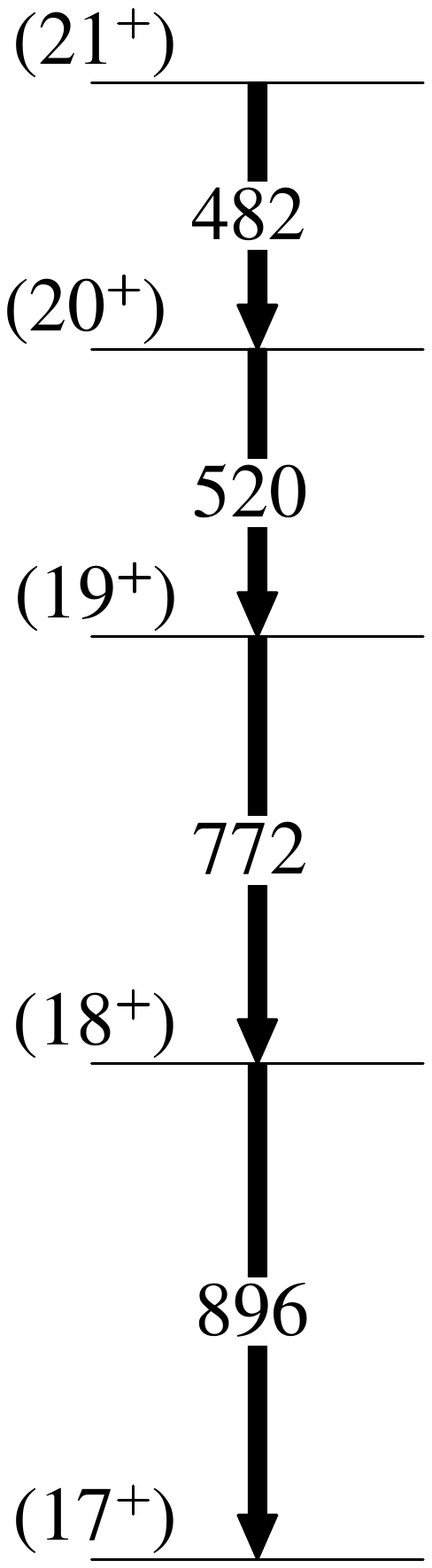}}
\put(1.1,1.2){\textbf{(a)}}
\put(6.5,1.2){\textbf{(b)}}
\end{picture}
\caption{\label{theo-cal7} (Color  online) The partial level scheme of the (a) $^{145}$Sm and (b) $^{146}$Eu nuclei showing the possible candidate of the attractive shears bands. The level scheme are adopted from the Refs. \cite{odhara, eidegu}.}
\end{figure}

\begin{figure}[t]
\centering
\setlength{\unitlength}{0.05\textwidth}
\begin{picture}(10,8.0)
\put(-0.6,-0.3){\includegraphics[width=0.59\textwidth, angle = 0]{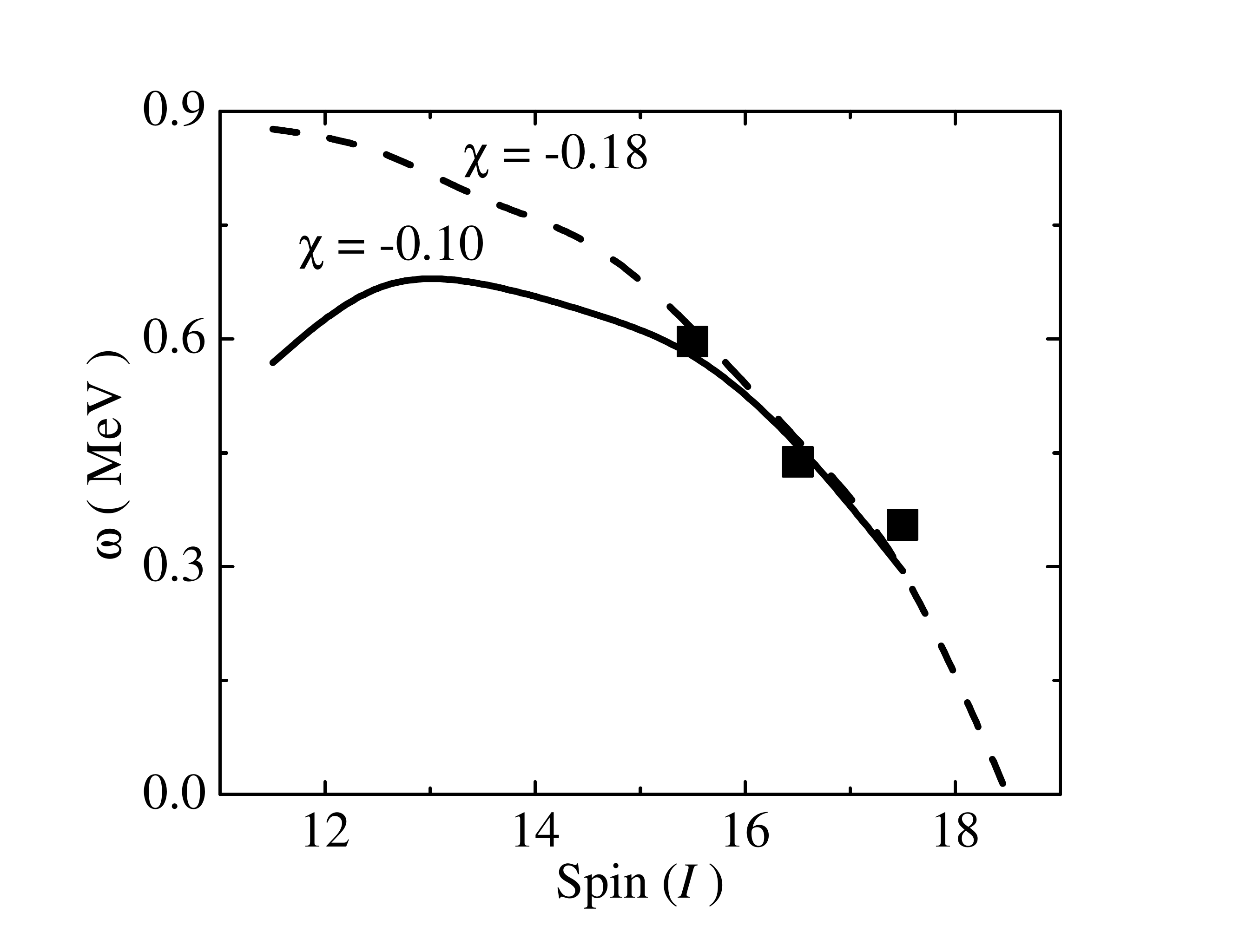}}
\put(+1.55, 1.2){\includegraphics[width=0.27\textwidth, angle = 0]{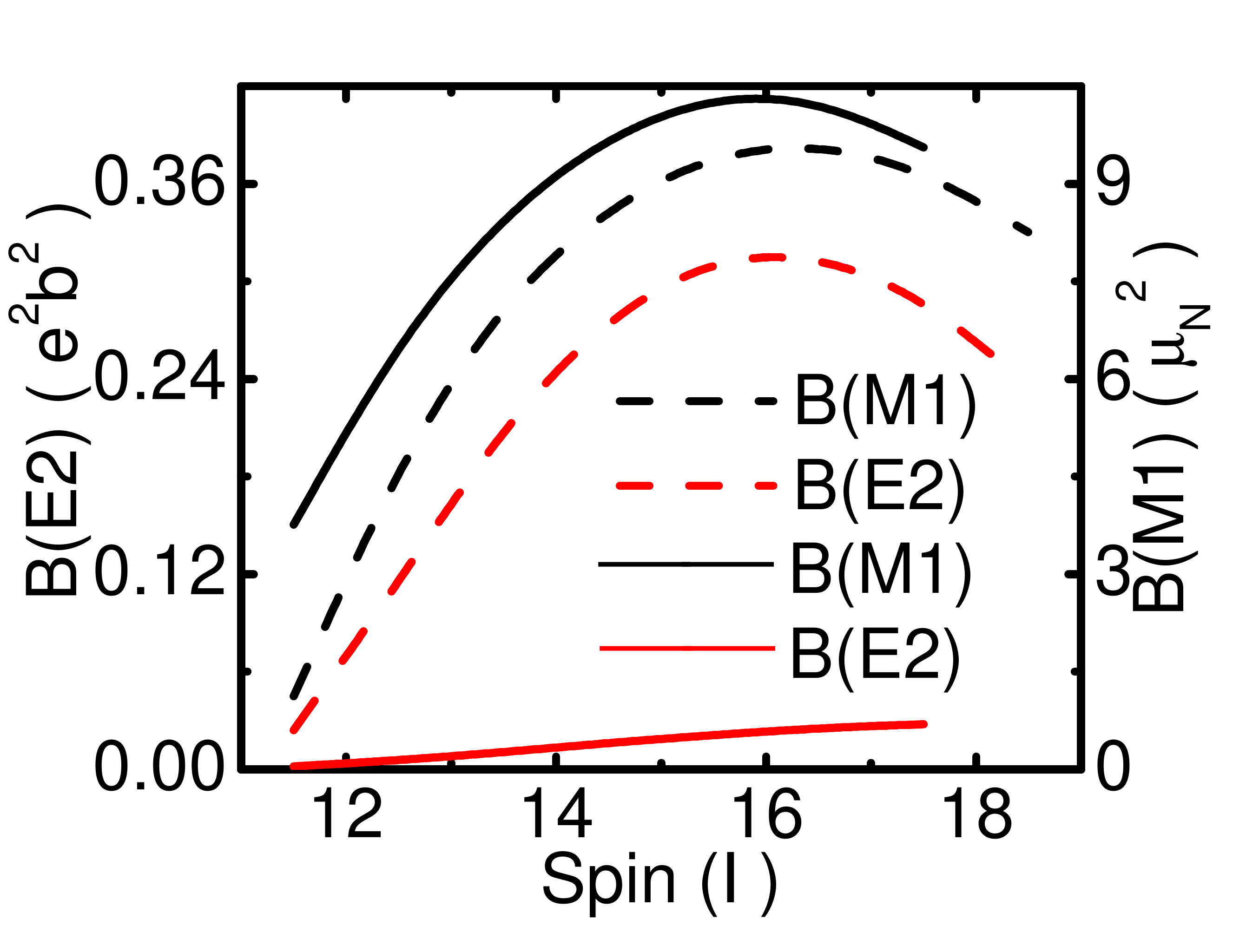}}
\end{picture}
\caption{\label{theo-cal-sm} (Color  online) The comparison of the experimental frequency with the calculated values within the HSPAC model considering the initial alignment of the vectors are anti-parallel to the axis of rotation (solid lines) and to the symmetry axis (dash lines) for the candidate of the attractive shears band in $^{145}$Sm. The calculated $B(M1)$ and $B(E2)$ values for the band are also shown in the inset.}
\end{figure}

As, the two modes of shears mechanism (MR and AMR) have been observed in the weakly deformed nuclei in $A$ $\sim$ 140 mass region, this new form would also be expected here for nuclei with $N$ $>$ 82. This is due to the fact that here, neutron(s) is(are) outside the $N$ = 82 core and protons can be easily excited to the $h_{11/2}$ orbital from the $g_{7/2}$/d$_{5/2}$ orbital make them to realize as particle character. In reality, the excitation spectra of the $^{145}$Sm \cite{odhara} and $^{146}$Eu \cite{eidegu} nuclei exhibit the band-like structure consisting of the strong M1 transitions having decreasing gamma energy with spin.

Above the 29/2$^{(+)}$ 4389-keV excited state of the $^{145}$Sm, a dipole band-like structure having M1 $\gamma$ transitions of energies 715, 477, 397 and 313-keV was observed by the A. Odahara $et$ $al$ \cite{odhara} (Fig. \ref{theo-cal6} (a)). The decreasing nature of the observed frequency of states hints the possibility of generation this band from the attractive shears mechanism. Though the parity of the states of interest are tentatively assigned the maximum observed spin of the band allows the possible configurations $\pi$$h_{11/2}^{2}$g$_{7/2}^{2}$ $\otimes$ $\nu$$i_{13/2}^{1}$ and $\pi$$h_{11/2}^{3}$g$_{7/2}^{1}$ $\otimes$ $\nu$$i_{13/2}^{1}$ assuming the parity of the states as positive and negative, respectively. Similar decreasing trend of the rotational frequency has been observed in the dipole sequence consisting of the $\gamma$-ray transitions of energies 896, 772, 520 and 482-keV built on the 4160-keV excited state in $^{146}$Eu \cite{eidegu} (Fig. \ref{theo-cal6} (b)). The observed maximum spin of this band can be reproduced from the configurations $\pi$$h_{11/2}^{3}$g$_{7/2}^{2}$ $\otimes$ $\nu$$i_{13/2}^{1}$ and $\pi$$h_{11/2}^{2}$g$_{7/2}^{3}$ $\otimes$ $\nu$$i_{13/2}^{1}$ assuming the negative and positive parity of the same, respectively. Though the neutron sector of the above configurations, both in $^{145}$Sm and $^{146}$Eu, have one neutron in $i_{13/2}$ orbital the possibility of occupying the $h_{9/2}$/$f_{7/2}$ cannot be ruled out. Such occupation of the neutron in the $h_{9/2}$/$f_{7/2}$ orbital would results change in the parity of the configuration but the maximum spin that can be generated within this framework (attractive shears mechanism) remains almost unaltered.

\begin{figure}[t]
\centering
\setlength{\unitlength}{0.05\textwidth}
\begin{picture}(10,8.0)
\put(-0.6,-0.3){\includegraphics[width=0.59\textwidth, angle = 0]{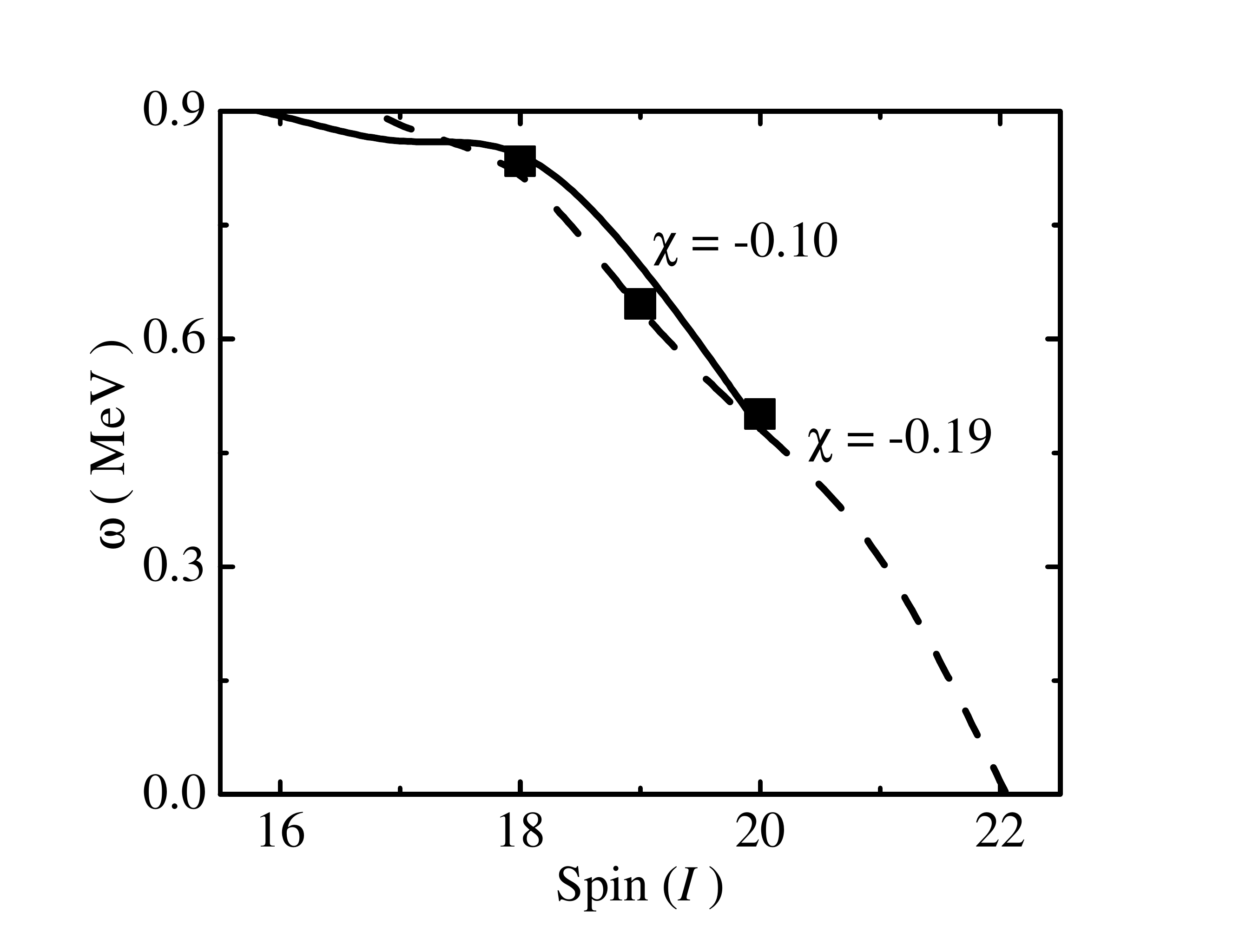}}
\put(+1.55, 1.25){\includegraphics[width=0.27\textwidth, angle = 0]{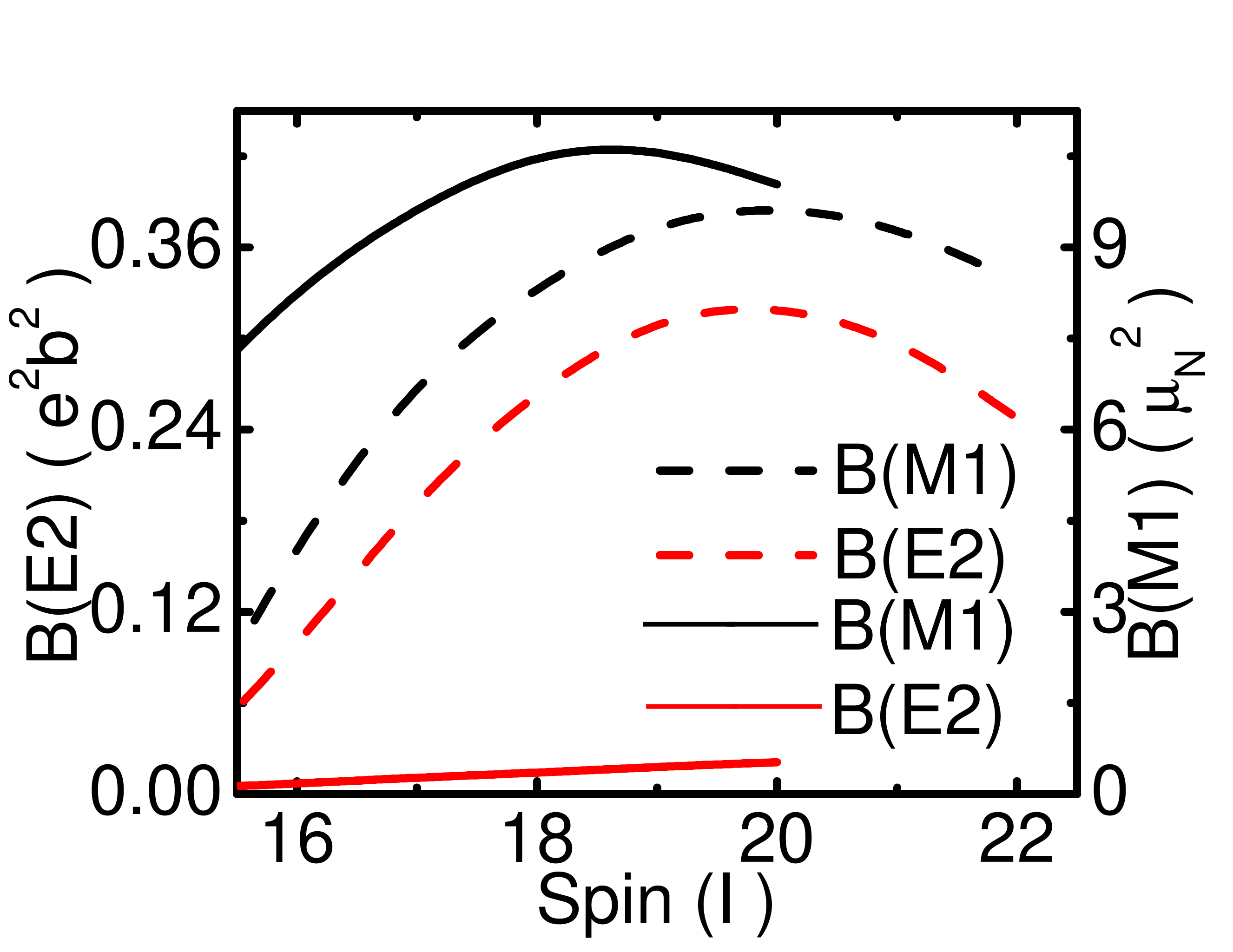}}
\end{picture}
\caption{\label{theo-cal-eu} (Color  online) The comparison of the experimental frequency with the calculated values within the HSPAC model considering the initial alignment of the vectors are anti-parallel to the axis of rotation (solid lines) and to the symmetry axis (dash lines) for the candidate of the attractive shears band in $^{146}$Eu. The calculated $B(M1)$ and $B(E2)$ values for the band are also shown in the inset.}
\end{figure}

To understand the characteristic of the band of interest, we have performed the HSPAC model calculations assuming the angular momentum vectors produced from the configuration of interest, as mentioned above, are anti-parallel to the axis of rotation and to the symmetry axis. In the calculations, we have considered the values of the parameters $Q_{eff}$, $Q_{coll}$, $g_{1}$ and $g_{2}$ are same as used in the calculations shown in Fig. \ref{theo-cal5}. For $^{145}$Sm nucleus, both of the calculations are well reproduced the experimental frequency considering the values of $\chi$, $j_{1}$ and a are -0.18, 6.5$\hbar$ and 2.62; and -0.10, 17$\hbar$ and 0.38 for the initial anti-parallel to the axis of rotation and to the symmetry axis, respectively (Fig. \ref{theo-cal-sm}). Here, for the initial anti-parallel to symmetry axis the adopted value of $j_{1}$ (17$\hbar$) can only possible for the negative parity configuration $\pi$$h_{11/2}^{3}$g$_{7/2}^{1}$ $\otimes$ $\nu$$i_{13/2}^{1}$. For the calculations of $^{146}$Eu nucleus the adopted values of $\chi$, $j_{1}$ and a are -0.19, 6.5$\hbar$ and 3.15; and -0.10, 19.5$\hbar$ and 0.33 for the initial anti-parallel to the axis of rotation and to the symmetry axis, respectively (Fig. \ref{theo-cal-sm}). The calculated frequencies in comparison with the experimental results (taken from Ref. \cite{eidegu}) are depicted in Fig. \ref{theo-cal-eu}. The above calculations reveal that the dipole bands of interest in $^{145}$Sm and $^{146}$Eu nuclei, as shown in Fig. \ref{theo-cal7}, may be originated due to the attractive shears mechanism.  

The present HSPAC calculations on the $^{145}$Sm and $^{146}$Eu nuclei reveal the increasing nature of the $B(M1)$ and $B(E2)$ transition strengths with spin unlike the decreasing nature of the transition strengths ($B(M1)$ and $B(E2)$ values) of the MR and AMR bands (inset of Figs. \ref{theo-cal6} and \ref{theo-cal7}). This intrinsic feature of these bands makes them unique in the group of the shears band. Therefore, to get a concrete evidence of the attractive shears mechanism the measurements of the transition strengths are utmost important. Thus measurement of the transition probability has been required on urgent basis to validate the strong candidature of such an extreme excitation mechanism in $^{145}$Sm and $^{146}$Eu nuclei.

\section{CONCLUSION}

It is well established the magnetic and antimagnetic rotational bands in the weakly deformed nuclei near shell closure are generated from the general mode excitation called shears mechanism. The present semi-classical calculations within the HSPAC model exhibit that the large collective contribution ($\chi$ = +0.54) of the MR in $^{139}$Sm (N = 77) is in agreement with the TAC and the SPAC results as performed in the previous work. On the other hand, least rotational contribution ($\chi$ = +0.13) results the maximal contribution of the angular momentum arises from the quasi-particle interaction for the MR band in $^{143}$Eu. All the observed MR bands in the $A$ $\sim$ 140 region are fall in between these two extreme values of $\chi$. Thus, it may be concluded that there exists an island in the values of $\chi$ as +0.10 $<$ $\chi$ $<$ +0.55 outside which no MR bands due to the repulsive interaction would be observed. 

The semi-classical calculations within the shears mechanism with the principal axis cranking model (HSPAC) reveal the possibility of the shears band due to the attractive interaction for nuclei in A $\sim$ 140 region. Though the limited information exists in literature, the decreasing trend of the experimental frequency in agreement with the HSPAC calculation exhibits the dipole bands above the 4389-keV and 4160-keV excited states in $^{145}$Sm and $^{146}$Eu nuclei demand their candidature of the attractive shears bands. Other interesting consequence of the HSPAC calculations is the increasing nature of the $B(M1)$ and $B(E2)$ transition strengths with spin unlike the decreasing nature of the transition strengths ($B(M1)$ and $B(E2)$ values) of the magnetic rotation and antimagnetic rotation. Thus measurement of the transition probability has been required on urgent basis to validate the strong candidature of such an extreme excitation mechanism in $^{145}$Sm and $^{146}$Eu nuclei.

\begin{center}
$\textbf{ACKNOWLEDGMENTS}$
\end{center}

The author would like to acknowledge the financial assistance from the University Grants Commission - Minor Research Project (No. PSW-249/15-16 (ERO)). The author would like to thank Professor A. Goswami for helpful discussion and suggestions during the hybrid shears mechanism with the principal axis cranking model (HSPAC) calculation.

\end{document}